\documentstyle[preprint,aps,epsf,floats]{revtex}
\tighten
\begin{document}

\preprint{\bf June 2000}
\title{Annual Report for the Department of Energy: 2000\\
Nuclear Theory Group\\
University of Washington}
\author{A. Bulgac, E. M. Henley, G. A. Miller, M.  J. Savage and L. Wilets
}
\address{
Department of Physics, University of Washington, 
\\
Seattle, WA 
98195.
}
\maketitle
\vskip 2in

\noindent This document is a summary of the physics research 
carried out by the Nuclear Theory Group at the University of Washington
during the last twelve-month period.

\vskip 0.5in

\noindent Current group members are
\begin{enumerate}
\item {\bf Faculty}: {\it Aurel Bulgac, Ernest Henley, 
Gerald Miller, Daniel Phillips,
Martin Savage\footnote{Joint Position with Jefferson Laboratory.}
 and Larry Wilets;}
\item {\bf Postdoctoral Fellows}: 
{\it Silas Beane, Christoph Hanhart and Michael Strickland;}
\item {\bf Graduate Students}: {\it Daniel Arndt, Jason Cooke, 
Gautam Rupak, Brian Tiburzi, and Yongle Yu;}
\item {\bf Administrative Assistant}: {\it Ahmbur Blue.}
\end{enumerate}

\vfill\eject
\tableofcontents
\vfill\eject

\section{Effective Field Theories in Multi-Nucleon Systems}

Since Weinberg's pioneering efforts to describe nuclear systems with 
effective field theory (EFT)~\cite{Wein}, 
nuclear physicists have spent the last decade 
attempting  to realize this dream.
Beane, 
Chen\footnote{PhD student of Savage who graduated in 1999 and is
currently a Postdoctoral Fellow at the University of Maryland}, 
Grie\ss hammer\footnote{A Postdoctoral Fellow in our group during 1997-1999,
and currently a researcher at Munich.}, 
Phillips\footnote{Currently a Research Assistant Professor in our group,
and who will assume a tenure-track  Assistant Professor position at Ohio
University in the fall of 2000.}, 
Rupak\footnote{ Currently a PhD student of Savage, who will assume a
Postdoctoral position at TRIUMF in the fall of 2000.}
and Savage 
at the University of Washington
have on-going programs to develop a systematic and 
perturbative theory to describe multi-nucleon systems.
During the last year, important progress has been made in our ability to 
calculate observables in the two- and three-nucleon sectors.
Techniques have been developed that enable precision calculations
of low-energy observables, particularly
those involving the deuteron.

\subsection{$np\rightarrow d\gamma$ for Big-Bang Nucleosynthesis}
\begin{figure}[!ht]
\centerline{{\epsfxsize=3.0in \epsfbox{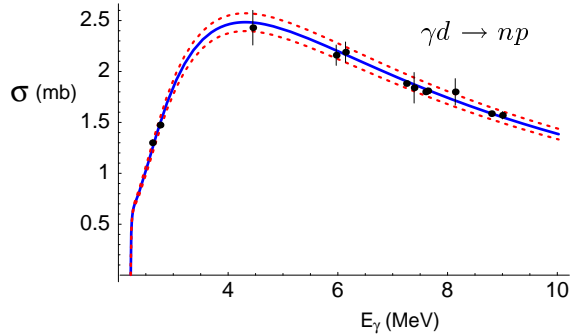}} }
\noindent
\caption{\it
The cross section for $\gamma d\rightarrow np$
versus photon energy. 
The solid curve corresponds to the theoretical cross 
section computed with EFT.
The two dotted curves correspond to
the $\sim 3\%$ uncertainty in the calculation of
Chen and Savage\protect\cite{CSa}.
Rupak's calculation reduced this uncertainty further, to below 
$\sim 1\%$\protect\cite{Ra}.
}
\label{fig:pdhigh}
\vskip .2in
\end{figure}

{\it J.-W. Chen, G. Rupak and M. J. Savage}
\vskip 0.1in
The recent revelation by 
Burles, Nollet, Truran and Turner
\cite{BNTTa} 
that the theoretical uncertainty in the cross section for 
$np\rightarrow d\gamma$ at low-energies 
(which they estimated to be $\sim 5\%$)
made a significant contribution to the 
uncertainties of elemental abundances computed in the Big-Bang, motivated 
Jiun-Wei Chen, Gautam Rupak and Martin Savage
to examine this process with EFT.
Chen and Savage\cite{CSa}
were able to derive expressions for the cross section that
had an associated error of $\sim 3\%$ over the energy region of interest.
This amplitude involves one constant that is not determined by
nucleon-nucleon scattering data, however,
it is determined by the precisely measured
cross section for thermal neutron capture.
The predicted cross section for 
$\gamma d\rightarrow np$ is shown in Figure~\ref{fig:pdhigh}.
%
%
Rupak\cite{Ra} was able to extend this calculation 
to one higher order
in the EFT expansion and so obtain a cross section 
with an associated uncertainty of $\sim 1\%$.
An additional constant  appears at this order and 
is determined by the cross section for $\gamma d\rightarrow np$
at higher energies.
Consequently, the theoretical uncertainty for this process 
that enters nucleosynthesis codes 
is reduced by a factor of $\sim 5$,  thereby eliminating this process
as a significant source of uncertainty.

\vfill\eject

\subsection{Photonic Interactions with the Deuteron in Chiral
Perturbation Theory}

{\it S.~R. Beane, M. Malheiro (Brazil), 
J. McGovern (Manchester),  D.~R. Phillips, }

{\it and 
U. van Kolck\footnote{Former Research Assistant 
Professor in our group
who will assume a tenure trach Assistant Professor position
at the University of Arizona in the fall of 2000.}
 (Caltech)
}
\vskip 0.1in

The electric and magnetic polarizabilities of the neutron and proton
provide great insight into the structure of the nucleon.
Due to the absence of pure neutron targets, efforts to extract 
neutron polarizabilities from processes involving nuclei continue.
The reactions 
\begin{eqnarray} 
\gamma + d &\rightarrow& \gamma + d \nonumber\\ 
\gamma + d &\rightarrow& \gamma + p + n 
\label{eq:reac} 
\end{eqnarray} 
are of particular interest because, provided suitable kinematics are
chosen, they may allow the extraction of electromagnetic
polarizabilities of the neutron.
In baryon chiral perturbation
theory hadronic effects can be included systematically, order-by-order
in the chiral expansion.  Consequently, the hadronic part of the
reactions (\ref{eq:reac}) can be calculated in a controlled way, and
effects due to electromagnetic polarizabilities of the neutron can be
reliably extracted from the data.  
\begin{figure}[!ht]
   \epsfysize=5.5cm
   \centerline{\epsfbox{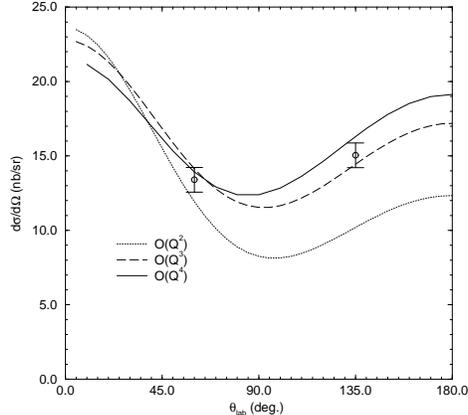}}
\noindent
\caption{
Deuteron Compton scattering in chiral perturbation theory as compared
to the data from Illinois at photon energy $69~{\rm MeV}$.
The dotted line is the leading-order result,
dot-dashed next-to-leading order, and the solid line is our
preliminary $O(Q^4)$ calculation.
  }
\label{fig:compa}
\end{figure}

We have recently computed the elastic photon-deuteron scattering to
next-to-leading order using Weinberg's power counting\cite{Wein}. 
The results
appeared in~\cite{Compton}, and are in reasonable agreement with the data
from Illinois~\cite{Lucas}, although they differ from the recent SAL 
data~\cite{Hornidge}---as do
all reasonable extant theoretical calculations. During the fall of 1999
Dr. Malheiro visited our group, and considerable
progress was made 
towards performing the calculation to next order. 
This higher-order calculation has the advantage that the neutron
polarizabilities appear as the only free parameter at this order in
chiral perturbation theory. 
It also allows for a test of the convergence
of the chiral expansion for this reaction. 
In Figure~\ref{fig:compa}
the results of three orders for this calculation are compared to the 
data at a photon energy of $69~{\rm MeV}$~\cite{Lucas}.

\vskip 0.1in
\subsection{Pion-Deuteron Scattering in Chiral Perturbation Theory}

{\it S.~R. Beane, 
U.-G. Mei\ss ner (Julich), 
and D.~R. Phillips}
\vskip 0.1in

The recent complete one-loop calculation of pion-nucleon scattering
in chiral perturbation theory~\cite{nadia}
has greatly improved the description of pion-nucleon
scattering.
In fact, the agreement with the low-energy 
pion-nucleon scattering data is now quite good. 
The logical next step is to apply the recently developed 
techniques to describe the deuteron in EFT
to pion-deuteron scattering. 
This is of particular interest
because the pion-deuteron scattering experiment constrains the
isoscalar combination of $\pi-N$ scattering lengths, $a^+$. 
However,
as in the case of the neutron polarizability, if one is
to extract this quantity reliably one must employ a formalism
in which the two-body corrections to the reaction mechanism
are under control. 
We have had preliminary discussions
regarding this calculation.

\vskip 0.1in

\subsection{Improving the Convergence of NN Effective Field Theory}

{\it D.~R. Phillips, G. Rupak, and M.~J. Savage}
\vskip 0.1in

At energies well below the pion mass the only relevant degrees of
freedom in few-hadron systems are the nucleons themselves. This led to
the formulation of an effective theory of $NN$ interactions which is
akin to effective range theory. 
Indeed, for two-nucleon processes this
effective field theory uniquely reproduces the results of effective range
theory. 
However, its predictions differ for processes in which
external probes couple to the $NN$ system, in that one can
systematically include the effect of operators which are not
constrained by $NN$ phase shifts, or related to $NN$ processes by current
conservation. 
There have been a number of very accurate calculations
pursued in the last year in this ``$NN$ EFT", but one persistent problem was
that the series for elastic processes on the deuteron converged
somewhat slowly.

Last summer we realized that this could be resolved by using an
alternative fitting procedure in the $NN$ EFT.
This procedure, called the Z-parameterization, involves reproducing
the tail of the deuteron wave function correctly in the $NN$ EFT
at next-to-leading order. 
For many observables this results in excellent convergence
of the expansion. 
Indeed, it means that in elastic scattering the
first corrections to the NLO result come from two-body operators and
relativistic effects.  
The Z-parameterization significantly improves the convergence of
inelastic deuteron processes, such as $np \rightarrow d \gamma$ 
as described previously, and
is no worse for ordinary $NN$ scattering than the original fitting
procedure\cite{CRSnopi}.

\vskip 0.1in
\subsection{Neutrino-Deuteron  Interactions}
{\it J.-W. Chen and 
M. N. Butler (Halifax)
}
\vskip 0.1in

Nuclear physics is playing a central role in the present discovery of
neutrino masses, the first experimental evidence for 
physics beyond that standard model.
The SNO detector will measure both the neutral and charged current interactions
of neutrinos from the sun thereby measuring, to some accuracy, 
their flavor composition 
(electron verses muon, tau or sterile).
Both the production of $pp$ neutrinos, and the subsequent flavor discrimination
involve inelastic deuteron processes.
Thus it is of the utmost importance to understand the electroweak interactions
of the deuteron.

\begin{figure}[!ht]
\epsfysize=1.6in
\centerline{\epsfbox{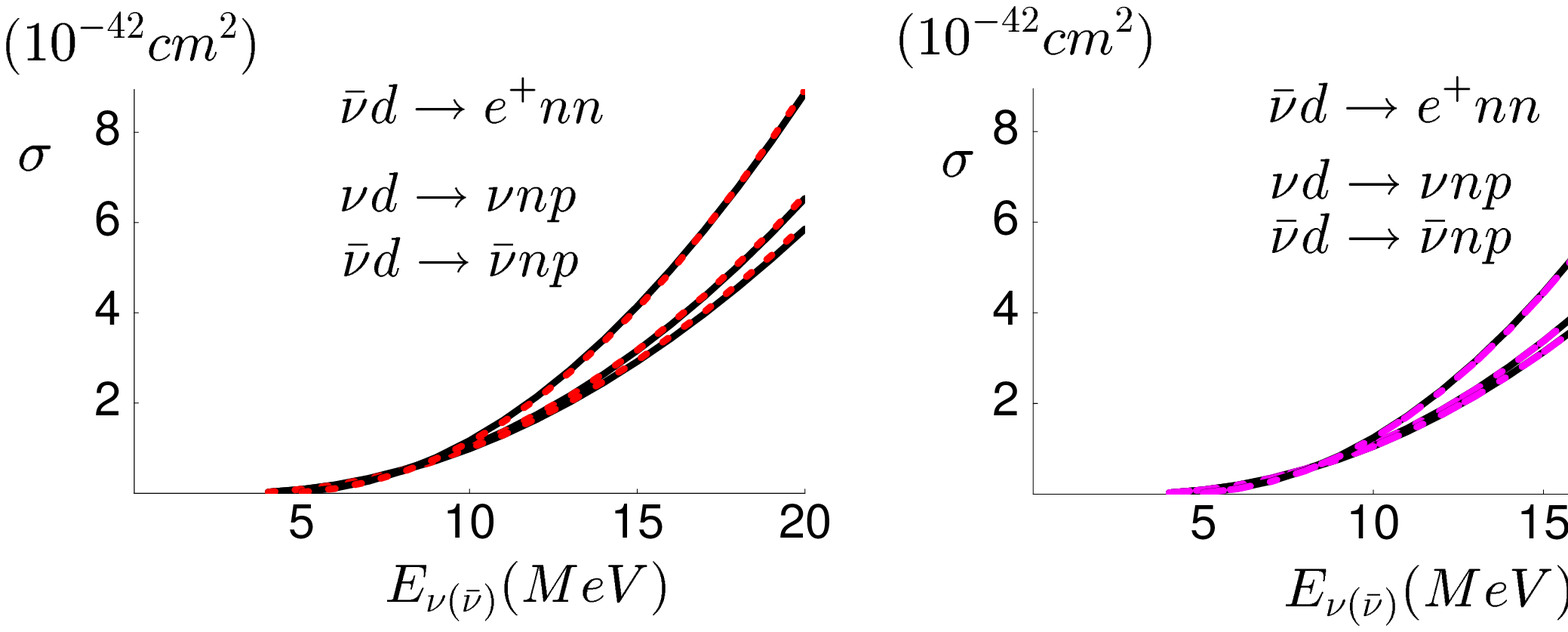}}
\caption{ 
Inelastic $\protect\nu (\bar{\protect\nu})d$ 
cross-sections versus incident 
$\protect\nu (\bar{\protect\nu})$ energy. 
The solid curves in the left graph are results of \protect\cite{KN}
while the dot-dashed curves,
which lie on top of the solid curves, are NLO in EFT with 
$L_{1,A}=6.3\ {\rm fm}^{3}$. 
The solid curves in the right graph are the results of \protect\cite{YHH}
while the dashed curves, which also lie  on top of the solid curves,
are NLO in EFT with $L_{1,A}=1.0\ {\rm fm}^{3}$.  }
\label{fig5}
\end{figure}
Existing potential model calculations\cite{YHH,KN}
of the $\nu-d$ breakup processes disagree 
at the $5\%$ level due to different assertions about meson exchange currents.
Effective field theory calculations by 
Butler and Chen\cite{BCa,BCb} of $\nu-d$ breakup processes have clarified this
ambiguity.
A local, gauge invariant operator contributes to the matrix element of the
axial
current at next-to-leading order, with coefficient $L_{1,A}$.
Different potential model results correspond to different choices of this
counterterm, as can be seen in Figure~\ref{fig5}.
This counterterm can be fixed in two possible ways.
Firstly, an experimental measurement of $\nu_e d\rightarrow ppe^-$ at
low-energies at the $1\%$-level, would determine the counterterm, enabling
$1\%$ predictions of the processes utilized by the SNO experiment (a higher
precision prediction would require knowledge of additional counterterms).
Secondly, a theoretical calculation of the rate of tritium $\beta$-decay would
also fix the counterterm.  This is being considered by Paulo Bedaque, a
Research Assistant Professor at the INT.
In fact, sophisticated potential model calculations\cite{PotSch} 
have implemented the later
approach to predict $pp\rightarrow d e^+ \nu_e$.
Butler and Chen have recently included the Coulomb corrections for 
$pp\rightarrow d e^+ \nu_e$.

\subsection{Three-Body Processes}
{\it P. F. Bedaque (INT), 
F. Gabbiani (Duke)
and H. Grie\ss hammer} 
\vskip 0.1in

During the last twelve months, Bedaque, Gabbiani and Grie\ss hammer
completed interesting work (that was underway while Grie\ss hammer
was a Postdoctoral Fellow in our group)
in three-body systems with effective field theory.
Phase shifts for $n-d$ scattering in the spin quartet channel and 
in several partial waves have been calculated both above and
below the deuteron breakup threshold, solely in terms of 
two-nucleon scattering  parameters\cite{BGGa}.
The agreement with data is impressive,  and agrees well with 
the few predictions 
that have been made using  sophisticated nuclear potentials.

\vskip 0.1in
\subsection{$NN \to NN\pi$ from the Effective Field Theory Point of View
and the Three Nucleon System}
{\it C. Hanhart, G. A. Miller and U. van Kolck (Caltech)}
\vskip 0.1in

For almost a decade high quality data available for the
near-threshold pion production in nucleon--nucleon collisions
has been available. However,
the reaction is still not fully understood. 
As pion dynamics are 
controlled by chiral symmetry constraints it was hoped that
chiral perturbation theory could resolve the uncertainties. Until 
recently, however, there was significant disagreement between
the chiral perturbation theory calculations and  data. Our work\cite{HMKa}
highlighted the reason for that: once a proper counting scheme is employed
it becomes clear that loops enter at next-to-leading order and
thus the tree level calculations carried out thus far were incomplete.
In addition we were able to demonstrate that p-wave pion
production is better behaved: loops enter only at next-to-next-to
leading order. 
The convergence of the chiral expansion was demonstrated
with a spin observable for $\pi^0$ production, where a parameter-free
prediction is in agreement with the data. At the order to which 
we worked
there were two unknown coefficients in the chiral Lagrangian.
The respective interaction terms not only influence observables 
in pion production but might provide a  solution to the  long standing
problem in few body physics, the so called $A_y$ puzzle in $pd$ scattering.
Thus our work demonstrated the close connection between pion production
and the three nucleon system. This  provides a deeper
understanding of low and medium energy nuclear physics.

\vfill\eject
\subsection  {Pion Production in Low Energy Two-Nucleon Interactions} 
{\it C. Hanhart, G. A. Miller, 
F. Myhrer (South Carolina), 
T. Sato (Osaka), }

{\it and 
U. van Kolck (Caltech)}
\vskip 0.1in

Chiral perturbation theory (or effective field theory)
is a relatively new way to derive nuclear forces and
cross sections in terms of parameters that carry information
about QCD dynamics.
Chiral power-counting arguments are expected to 
supply an organizing principle.

Understanding 
  threshold  pion  production reactions
in nucleon-nucleon  collisions provides an interesting challenge to any such
approach
because the relevant momentum scale is $p\sim\sqrt{m_N m_\pi}$, so
$p\over m_N$
is not very small and the usual low momentum expansion is slowly convergent.
The acquisition 
of excellent data\cite{38} for the $p p \rightarrow p p \pi^0 \;$
reaction heightened the interest in studying these reactions.
Our first paper on this  subject
found that the seemingly  mandated (by chiral symmetry) treatment of the
off-shell pion-nucleon scattering amplitude caused the
term arising from the pion rescattering to interfere destructively with
other amplitudes, leading to a
computed cross section was considerably smaller
than the measured one\cite{39}.
To be specific, consider the effect of a term
$\dot{\pi}^2 \bar N N$. 
Evaluation of this term, for threshold kinematics,
in the rescattering graph
(without including initial or final state interactions
between the nucleons) gives a factor of $m_\pi m_\pi/2$, in which the first
factor arises from the final energy of the pion and the second from
the energy of the intermediate pion which carries half of the initial
center of mass energy.
But another group,
Sato et al.\cite{41}  treated this same factor as
$m_\pi(E(p)-E(p'))$ 
in which $p,(p')$ is
the relative momentum in the initial (final) state, and
$E(p)=\sqrt{p^2+m_n^2}$. This is evaluated
by multiplying by the initial and final state $pp$ wave functions and
integrating 
over the momenta. Their result is  that
the low momentum tail of the initial high energy $pp$ wave function
and the high momentum tail of the final threshold wave function turn out to
be dominant, so that 
this difference in energies turns out on average to be much larger than
$m_\pi$ in magnitude and
negative. Our discussions at the joint INT/ANL August 1998 workshop
led us to try to understand better
the  differences between the calculations.
Both methods of calculation  employ a three-dimensional
integral to the evaluation of a Feynman diagram  which is  a
four-dimensional integral. This led us  to originate a semi-realistic
toy model in which
one may evaluate the four-dimensional
integral exactly, and determine which (if
any) of
the two approximations are accurate.
We proposed to evaluate the relevant
Feynman graphs.
The preliminary results of the toy model calculations are that the procedure
of  Ref.~\cite{39} is a good approximation to the exact toy model
answer.

\vskip 0.1in
\subsection {Effective Field Theory for $\bar {N}N$ Scattering}

{\it M. Alberg\footnote{Former PhD student of Wilets and 
currently a faculty member at Seattle University.} 
(Seattle U.), E. M. Henley}
\vskip 0.1in

An effective field theory for nucleon-anti-nucleon low energy scattering is being
developed. The aim is obtain phase shifts which can better fit the scattering
data, so as to extend previous work on various nucleon-anti-nucleon interactions.
The case of a large effective range and a small scattering length has been
examined.

\vskip 0.2in

\section {Light-Front Nuclear Physics}

\vskip 0.1in
The aim is
to develop a relativistic treatment of nucleons in nuclei which can be
used in  a variety of high-energy, high-momentum transfer processes,
and which incorporates fully
the present knowledge of nuclear physics.
A method that yields covariant results is
needed to 
properly interpret a 
host of new electron-nuclear scattering experiments  performed
at  Jefferson Laboratory
 and at HERA in Hamburg. 
 Relevant reactions include 
deep inelastic lepton-nucleus scattering, the related
Drell-Yan ($\mu^+\mu^-$ production)  experiments, and 
 nuclear quasi-elastic reactions such as  $(e,e'),\; (e,e'p),\; $
and $(p,2p).$
Our choice has been to use
light-front variables. For example, in the
 parton model    the Bjorken variable $x$ is the
 ratio $x=k^+/p^+$ where  $k^+=k^0 +k^3$ is the plus-momentum of the struck
 quark and $p^+$ is the plus-momentum of the target. Quark distributions
 represent the probability that a quark has a plus-momentum fraction $x$.
 One of our main interests   is
  in computing the distribution functions $f(k^+)$ which give the probability
for nuclear nucleons and mesons to have a  given  plus-momentum $k^+$.

In the light-front technique the $f(k^+)$ are determined by the ground-state
wave function, while in
 the usual equal-time formulation obtaining the $f(k^+)$
 requires computing the response function  which involves matrix elements
between the ground and all excited states. Thus the use of the light
front dynamics engenders vast simplicities, if one is able 
to compute  the ground-state wave function using those  light-front
dynamics.
Our
previous work included substantial  progress towards computing realistic wave
functions, which is described below.
The light-front
 quantization for a chiral
  Lagrangian is performed by obtaining the appropriate Hamiltonian
and energy momentum tensor\cite{24,25}.
 The mean field approximation was defined and applied
to infinite, isospin symmetric, nuclear matter\cite{24}.
 The one boson exchange 
treatment of nucleon-nucleon scattering was studied and a close connection with
the on-shell T-matrix of the usual equal time formalism is obtained\cite{24}. 
 A new light-front one-boson exchange nucleon-nucleon potential,
which
achieves a reasonably good representation of the phase shifts
has been obtained\cite{25}.
 The tree-level pion-nucleon scattering scattering amplitude
 was shown to reproduce
  soft pion theorems\cite{24}
 Meson distribution functions for finite nuclei have been obtained,
using a simple model\cite{26}.
 Calculations of properties of finite nuclei are well underway.
One of the problems in using the light-front formalism
is the lack of manifest
rotational invariance. We have recovered the $2j+1$ single-nucleon degeneracy
characteristic of the shell model of spherical nuclei. This is accomplished by
deriving and solving a new light-front  equation for the
nucleon wave functions\cite{27}.
Nucleon-nucleon correlations are often treated using Brueckner
theory. A light
front version of this has been derived and successfully applied
to infinite nuclear matter. The simplicity of the vacuum allows us to
derive the theory from a set of integral equations which maintains the
necessary connection between the  mesonic components of the Fock space
and the nucleon-nucleon potential\cite{25}.
Good saturation properties are obtained\cite{28},
including a compressibility of 180 MeV. The nuclear medium is found to
enhance the average number 
of pions per nucleon by about $5\%$.

\vskip 0.2in 
\subsection{Heavy Nuclei}

{\it P. Blunden (Winnipeg), 
M. Burkardt\footnote{Former Research Assistant Professor at the UW.}
(New Mexico), 
G. Krein\footnote{Former Postdoctoral Fellow in our group and 
long-term visitor during the past academic year.}
(IFT),}

{\it R. Machleidt (Idaho), G. A. Miller}

\vskip 0.1in 

We have made progress in
using light-front dynamics to compute  nuclear wave functions.
The implementation of rotational invariance should ultimately
allow a widespread  use of light-front dynamics in any calculation
in which relativistic aspects of nuclei are expected to be relevant.
A necessary condition for this to occur is 
that the computed momentum distributions
$f(k^+)$ be at least roughly  consistent with the  experimental data
for deep inelastic lepton-nucleus scattering, as well that of the
$(e,e'p)$ reaction.
It is interesting to note that achieving even a vague resemblance
to the data
requires a very high level of nuclear structure theory. To explain this,
we review the 
history of  this problem. 
The   mean field calculation for infinite nuclear matter\cite{25} found that the
nucleons carry only about $65\%$ of the nuclear plus-momentum, with the
remainder carried by the nucleon. This fraction  should  be   about $90\%$
in infinite  nuclear matter\cite{42}.

These results  are  specific to the
use of  mean field theory 
and infinite nuclear matter. Thus it was necessary to make 
calculations for finite nuclei and to go beyond the mean field approximation.
Our  finite nucleus result\cite{27}, obtained using the mean field approximation
for $^{40}$Ca, is that the nucleons 
 carry only     $73\%$ of the plus momentum, with the $\omega$
carrying almost all of the remainder. Thus a substantial problem remains.

We  have done calculations 
beyond mean field theory by   including  correlations
between two nucleons. This is  a light-front version of Bruckner theory.
Our calculations\cite{25,28}
reproduce the saturation properties, but
including correlations
allows much weaker scalar and
vector potentials. 
One consequence is that the nucleons carry at least  $84\%$ of the plus momentum,
which is much closer to the desired value of $90\%$.

\begin{figure}
\unitlength1.1cm
\begin{picture}(8,7)(-2,-8.5)
\includegraphics{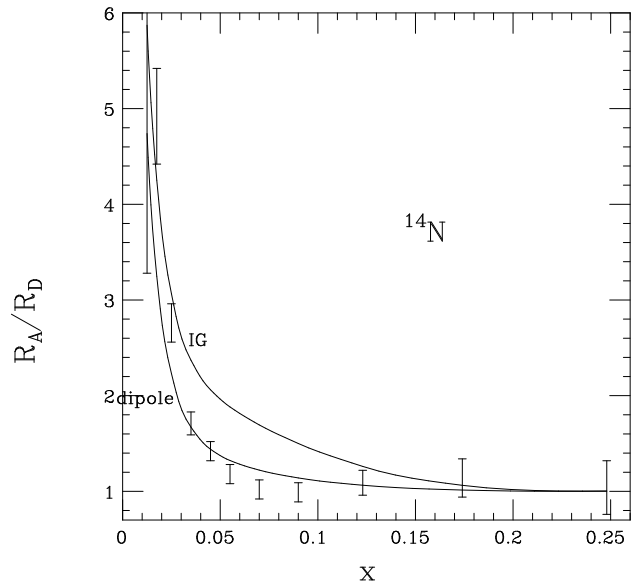}
\end{picture}
\caption{Ratio of $R={\sigma_L\over \sigma_T}$, R(A)/R(D), A=14.
 IG and dipole refer to the form factor used.}
\label{fig:newra}
\end{figure}

\vskip 0.1in
\subsection{HERMES Effect}

{\it S. J. Brodsky(SLAC), 
M. Karliner(Tel Aviv), and G. A. Miller}
\vskip 0.1in

We have emphasized the need to reduce the computed value of the
plus-momentum carried by the  $\omega$
mesons. But it is clear that  any future versions of the present
calculations will predict some 
significant enhancement of the nuclear $\omega$ content. The proposal stated
that ``We shall  try to find
 experimental signals for this presence.
This may be difficult, but it seems worthwhile to try.'' In June 1999 the
HERMES collaboration announced \cite{Ackerstaff:2000ac}
that the ratio $R=\sigma_L/\sigma_T$
was enhanced in nuclear deep inelastic scattering. A small value of $R$
is a signature that the struck partons have spin 1/2 \cite{Callan:1969uq},
so a very large value indicates the strong presence of nuclear mesons.

This HERMES discovery led us to search for the influence of vector
and scalar mesons. We 
showed \cite{Miller:2000ta}
that nuclear $\sigma$, and $\omega$,  mesons can contribute coherently to
       enhance the electroproduction
cross section on nuclei for longitudinal virtual photons
       at low $Q^2$
while depleting the cross section for transverse photons. We  described
recent HERMES inelastic lepton-nucleus scattering data at low $Q^2$ and
       small x using photon-meson and meson-nucleus couplings which
are consistent with
      (but not determined by) existing constraints from meson decay widths, nuclear
       structure, deep inelastic scattering,
and lepton pair
production data. We find that while
        pion currents
are not important for the present data, they could be
       observed at different kinematics.
Thus our model is very easy to test and is being tested by an experiment at
Jefferson Lab. The results of our calculation for $^{14}N$ are shown in 
Figure~\ref{fig:newra}.

\vskip 0.1in
\subsection{Relativistic Corrections to Nuclear Charge Radii}

{\it P. Blunden (Winnipeg), 
M. Burkardt (New Mexico), and G. A. Miller}
\vskip 0.1in

R. Michaels and P.A. Souder (of the TJNAF HAPPEX collaboration) have  
proposed\cite{RMa}
an  experiment (see also \cite{HPSM})
intended  to measure the neutron density
of $^{208}$Pb using parity violating electron scattering. The stated aim
is to determine  the neutron radius  to
about 1\%. This made it relevant to pursue a new study of
how 
relativistic effects influence the computed charge radii.
We showed that these effects are very tiny\cite{Blunden:2000wb,review}.

\vskip 0.1in

\subsection{Relativistic Quarks in Relativistic Nucleons}

{\it G. Krein (IFT), and G. A. Miller}
\vskip 0.1in

There has also been progress in going beyond
the nucleon-meson version of nuclear physics. 
There may be some processes such as
DIS for which quark models can be used directly. Thus our aim
is to provide a light-front calculation in which the nucleons are treated as
moving relativistically within the nucleus, and in 
which the the quarks (and gluons) are moving relativistically within the
nucleon. An extension of the quark-meson coupling model\cite{QMC}
may provide the easiest path to obtaining a realistic
formalism that is also tractable.
In this model, the quarks are assumed to be
bound in non-overlapping nucleon bags, and the interaction
between nucleons arises from a coupling of meson fields to the
quarks. First calculations using a quark-di-quark model of the nucleon
which is immersed in the medium give promising results. Nuclear
saturation is achieved with relatively weak mesonic fields.

\vskip 0.1in

\subsection{  Light Nuclei} 
{\it J. Cooke\footnote{PhD student of Miller.}
,  G. A. Miller, and D. R. Phillips}
\vskip 0.1in
This is the thesis project of  Jason  Cooke, which is aimed at doing
relativistic deuteron physics. We started  by examining the  bound states of 
 the Wick-Cutkosky model
in which  two-heavy scalar particles (``nucleons'')  interact by
exchanging another scalar particle.
This is a good testing ground for such calculations, because essentially
exact
solutions exist\cite{46}.
As a starting point
we  studied  bound states\cite{Cooke:2000yi} using
the Hamiltonian formalism on the light-front. In this approach
manifest rotational invariance is broken when the Fock space is truncated. By
       considering an
effective Hamiltonian that takes into account two meson exchanges,
       we find
that this breaking of rotational invariance is decreased from that which occurs
       when only one meson exchange
is included. The best improvement occurs when the
       states are weakly bound.

We also made a more detailed study\cite{Cooke:2000ef}
of the Wick-Cutkosky model in which
light-front potentials for two-nucleon bound
       states are calculated using two approaches.
First, light-front time-ordered
       perturbation
theory is used to calculate one- and two-meson-exchange potentials.
These potentials give results that
agree well with the ladder and ladder plus crossed
       box Bethe-Salpeter spectra. Secondly, approximations that incorporate
       non-perturbative
physics are used to calculate alternative one-meson-exchange
     potentials.
These non-perturbative potentials give better agreement with the spectra of
  the
full non-perturbative
ground-state calculation than the perturbative potentials. For
lightly-bound states, all of the approaches appear to agree with each other.

\medskip
The above results indicate that the detailed Fock-space wave function
of the  deuteron
could be obtained using light-front techniques.
The deuteron is the nucleus that has been the subject of almost all light
front
calculations of nuclei, and will be studied intensively at Jefferson Lab.
The previously mentioned papers can
be considered as a warm-up for a realistic calculation, which includes meson
degrees of freedom of  the deuteron.

\vfill\eject

\subsection{Nucleon Structure Functions}
{\it M. Alberg
(Seattle U.), E. M. Henley, and G. A. Miller}
\vskip 0.1in

The  nuclear calculations, mentioned above, which predict  a 
significant  $\omega$
content of the nucleus lead us to
consider the effects of the $\omega$ content of the nucleon. Recent Drell-Yan 
studies\cite{50}
of the proton and deuteron have allowed the extraction of the
$\bar d(x) / \bar u(x)$, and  $\bar d(x) - \bar u(x)$, 
which is equivalent to the
isovector and isoscalar anti-quark distributions of the nucleon.
Calculations using the effects of a pion cloud seem to successfully reproduce
the isovector $\bar q$ distributions$\cite{50}$,
but provide   isoscalar distributions that
are smaller than the observed ones. The proposal
said  ``we expect that including the
effects of the $\omega$ should supply the missing strength,
and propose to make
the necessary calculations. This subject is related to a project of
Henley \& Alberg
and we plan to join forces.'' The result \cite{Alberg:2000bc} was that
we were able to 
use the meson cloud model
of the nucleon to calculate distribution functions for
$(\bar {d} - \bar{u})$ and
$ \bar{d}/\bar{u}$ in the proton.
Including the effect of the
omega meson cloud, with a coupling constant $g_\omega^2/4\pi\approx 8$, allows
a reasonably good description of the data. 
We are also trying to understand how the ratio $\bar d(x) / \bar u(x)$
can fall below unity at large values of $x$. The only effects that can
give such a result involve the breaking of charge symmetry.

\vskip 0.1in
\subsection{Quantization of Spin $\frac{3}{2}$ 
Particles on the Light Front}
{\it C. Hanhart and G. A. Miller}
\vskip 0.1in

Up to now there is no consistent procedure available to quantize
higher-spin baryons on the light-front. However, quantization is a 
necessary step if light-front dynamics is to be used for nuclear applications.
So far we are able to quantize the free spin-$\frac{3}{2}$ field on the
light-front.

\vskip 0.2in
\section {
Color Transparency and 
High Momentum Transfer Reactions}
{\it  M. Frankfurt (Tel Aviv), G. A. Miller, 
M. Sargsian\footnote{Former Postdoctoral Fellow in our group and
currently an Assistant Professor at Florida International University.}
 (FIU), and 
M. Strikman (Penn. State)}
\vskip 0.1in

Color transparency\cite{29}  is the unusual
vanishing of initial and final state interactions in high
momentum transfer ($Q^2$) nuclear  reactions in which the
resolution is good enough to ensure that no extra pions are
created in the  fundamental hadronic two-body reaction. Examples
 are the  $(e,e'p)$  $(p,pp)$ and $(e,e'\Delta)$
reactions involving nuclear targets. Several experiments 
searching for color transparency  are
underway.
Much of the theory of the three fundamental aspects of color
transparency  has been  done 
 as well as possible, any future progress in the theory rests on the
 ability to interpret 
the findings of future experiments.

Our main recent progress  is in understanding 
coherent pion  diffraction into minijets.
Consider the  process of
 coherent disintegration of a high energy (500 GeV)
pion beam into two jets on a nuclear target. We
found\cite{15} in 1993 that if each of the jets
has a very large transverse momentum $\kappa_t> 4 $ GeV/c, but
the sum of  their momentum
is very close to that of the beam, the forward cross section is
predicted to vary with nucleon number $A$ approximately
as $A^2$,
 a very striking
prediction of color transparency.
 If one integrates 
the angular distribution to obtain the cross section
for this coherent process, the power of 2 is changed
to 4/3. 
We encouraged experimentalists  to search for this process, and
the  challenge was taken up
by the Tel Aviv group as part of Fermilab E791. They compared data taken
on $^{195}$Pt and $^{12}$C targets and found\cite{52}
a power of 1.55$\pm$0.05 in
rough but striking  agreement (in the sense that the ratio of the
cross sections for Pt and C targets was about 70 times bigger than the
prediction of the usual diffractive treatment) 
with our prediction.

We  proposed    to
use  perturbative quantum chromodynamics to input the perturbative part of the
pion wave function into the calculation, to
include  the final state $q\bar q$ interaction 
and, to  specify the quantum numbers of the different  final two-jet states.
We intend also to use an updated version of the
soft $q\bar q$-nuclear final state interaction and to include the effects of
the skewed gluonic distribution of the nucleon and
the effects of electromagnetic disintegration of the pion. Much of the work
has been done and some appears in a paper accepted for
publication\cite{Frankfurt:1999tq}. We are preparing a more detailed
calculation.

\vskip 0.1in

\subsection{ The High Energy $ \gamma d\to np$ Reaction}
\vskip 0.1in

We  have studied $\gamma d\to np$, which has been the
subject of a recent TJNAF experiment\cite{53}. 
A new  mechanism is introduced in which 
photon absorption by a quark in one nucleon followed by a high momentum
transfer interaction with  a quark in the other may produce two nucleons
with high relative momentum. In this mechanism the
amplitude
depends on the 
 well-known,  low-momentum, two-nucleon component of the deuteron
wave function. Other diagrams  are smaller by orders of magnitude.
We intend to determine how much of the observed
cross section can be accounted for in this way. We study
the  absolute value, energy, and
angular dependence of the recently measured cross
sections. According to the literature, 
two competing mechanisms account for the
high momentum transfer nucleon-nucleon
scattering amplitude. These are
the Feynman mechanism and the use of the minimal
Fock space components. We  studied the role of  each of these mechanisms
in determining the  $np$ final state interaction that occurs in the
photoabsorption process. It is possible that the careful analysis of this data
could lead to the resolution of a long-standing question about which mechanism
is most important. Our first calculation
is now published\cite{Frankfurt:2000ik}.

\vfill\eject

\section{Astrophysics}
\vskip 0.1in
\subsection{Neutrino and Axion 
Radiation from Neutron-Star
Cooling and Supernovae}

{\it C. Hanhart, D.~R. Phillips, and S. Reddy (INT)}
\vskip 0.1in

Neutrino-nucleon processes are important sources of cooling and
heating in several astrophysical phenomena. Neutrino cooling in
neutron stars, for example, is driven by reactions of the type
$nn\rightarrow nn\nu\bar{\nu}$ and $nn\rightarrow npe^-\bar{\nu}_e$.
To date, the strong interaction correlations between
the two nucleons have been calculated only perturbatively, i.e. using
the one-pion-exchange potential as the $NN$ amplitude~\cite{MF}. 

Based on soft-radiation theorems we performed a model-independent
calculation of the emission rates for $NN \rightarrow NN \nu
\bar{\nu}$ and $NN \rightarrow NNa$, finding that the
one-pion-exchange approximation overestimated these rates by a factor
of 4-5. This result will have a serious impact on the role played by
neutrinos in the cooling of neutron stars. It may also affect the
understanding of the role of neutrinos in supernova explosions. At
this stage it is unclear how this  result will 
change the axion mass bound, since a different neutrino rate will
influence those as well. However, it is clear that previous estimates
of the axion production rate in SN1987A were based on a two-body axion
emission amplitude that was much too large.

\vskip 0.1in
\subsection{SN1987A Bounds on Large Compact Extra Dimensions}

{\it C. Hanhart, D.~R. Phillips, S. Reddy (INT), 
and M.~J. Savage}
\vskip 0.1in

A scenario that is emerging from string theory and effective field theory
descriptions of gravity is that matter fields exist on a four-dimensional brane
embedded in a higher dimensional space which gravity alone can propagate.
One of the strongest constraints on the existence of large, compact,
``gravity-only", extra dimensions comes from SN1987A. If the rate of
energy loss into these putative extra dimensions is too large, then
the neutrino pulse from the supernova will differ from that actually
seen. The dominant mechanism for graviton and dilaton production in
the supernova is via gravistrahlung and dilastrahlung from the
nucleon-nucleon system. Low-energy theorems of the type discussed in
the previous paragraph can also be used to calculate these
processes in a model-independent fashion. This relates these
emissivities to the measured nucleon-nucleon cross section.
This is possible because for soft gravitons and dilatons the leading
contribution to the energy-loss rate is from graphs in which the
gravitational radiation is produced from external nucleon
legs. We have re-evaluated the
bounds on extra dimensions using our low-energy theorem, and find that
if there are two extra ``gravity-only" dimensions then to be
consistent with the SN1987A observations these dimensions must be of a
radius less than 1 micron.

\vskip 0.1in
\subsection{Shielding of Nuclei in a Finite 
Temperature Plasma}
{\it  B. Giraud (Saclay), 
J. J. Rehr,  M. J. Watrous\footnote{Former PhD student of Wilets.}
, and L. Wilets.}
\vskip 0.1in

A novel technique for evaluating the electron shielding of nuclei
in a finite temperature plasma has been developed and applied.  The physics 
is Hartee plus local-density approximation to account for exchange and 
correlation (Kohn-Sham-Mermin).
This is particularly applicable to the fusion process in stellar and 
laboratory plasmas.  Finit-temperature, finite-density electron
Green's functions are calculated.  Contour integration yields a sum of
residues over the complex Matsubara poles.  This automatically includes
discrete bound states and continuum states weighted by the finite 
temperature Fermi factor. The one--center problem is adequate for the 
calculation of fusion rates. A paper 
describing the method and presenting numerical comparisons with the 
works of others has been published\cite{WWRa}.

A second paper 
including shielding due to ions 
and applying the method to the solar fusion problem has been 
published\cite{WGWRa}.
We believe the results obtained are the most 
reliable reported to date.

\vskip 0.2in
\section{QCD at High Density}
{\it S. R. Beane, P. F. Bedaque (INT), and M. J. Savage}
\vskip 0.1in

The realization that high-baryon density systems can be 
color superconductors\cite{alford1}
has lead to significant progress during the past year or so.
In such dense systems perturbative calculations of their properties
in $\alpha_2 (\mu)$ are possible.
The very low-energy excitations of the dense system can be described with an
effective field theory of the pseudo-Goldstone bosons associated with the 
spontaneous breaking of the global symmetries of the underlying theory.
Previous works had attempted to determine the decay constants and 
masses of the pseudo-Goldstone modes, but contained errors.
We determined the decay constants and
masses of the excitations\cite{BBSa}, making extensive use of effective field
theory techniques.

The original analytical determinations of the superconducting gap in
QCD suffer from two flaws. First, the gap equation is divergent and so
must be regularized and renormalized.  Most treatments have taken the
baryon density as a sharp cutoff. This might cause some concern since
it leads to the possibility of contaminating the low energy physics of
the gap with ad hoc high energy physics.  Second, the solution of the
gap equation at momenta large compared to the gap has been obtained by
assuming a particularly unhealthy mathematical approximation.  In a recent
paper, we use cutoff regularization to define a renormalized gap equation. 
We
also find an {\it exact} asymptotic solution to this equation, thus
excising the flaws contained in previous determinations. Our results
confirm the original analysis by Son. 
The renormalized gap equation 
enabled us to resum the large logarithmic 
contributions due 
to the evolution of the strong coupling constant between $\mu$ 
and the gluon magnetic  mass scale~\cite{BBSb}.
Recently, Beane and Bedaque have derived an expression for 
the gap with dimensional regularization~\cite{BBc}.

\vfill\eject

\section{Fundamental  Symmetries }

Fundamental symmetries has long been a topic of high interest at the U.W.
Recent efforts involve the use of nuclear physics to elucidate the 
standard model.

\vskip 0.1in 

\subsection{Charge Symmetry Breaking}
{\it  G. A. Miller, J.A. Niskanen (Helsinki), 
U. van Kolck (Caltech)}

\vskip 0.1in

Charge symmetry, CS, is a fundamental symmetry which occurs
if the Lagrangian is invariant with respect to  rotating the u
into d quarks (or d$\to $ u) quarks\cite{35}. 
We were able to   
use chiral effective field theory to make predictions, based on QCD,
for the reaction $np\to d\pi^0$. If charge symmetry holds, the cross section is
symmetric about 90$^\circ$ in the cm frame. The form of the
charge symmetry breaking
part of the hadronic Lagrangian is determined by  the symmetries of
QCD\cite{ivwei1}.
This  Weinberg term predicts a large breaking of CS in the $\pi$N
scattering amplitude which  is needed to compute the cross section for
$np\to d\pi^0$. 
The numerical result was that
this Weinberg term  provides
the  dominant 
contribution to the forward-backward 
asymmetry in the
 angular distribution for  
the reaction $pn\to d\pi^0$, for reasonable
 values of the mass difference between down  and up quarks, 
$\delta m_N$.  Using a value $\delta m_N\approx$ 3 MeV leads
to a prediction of a 10 standard deviation effect for a TRIUMF
 experiment.

\vskip 0.1in

\subsection{Parity Violating Proton-Proton Scattering at High Energy}

{\it G. A. Miller}
\vskip 0.1in

Lockyer {\it et al.}\cite{56} found a very large
parity violating asymmetry $A_L$ of about
$A_L=2.5\times  10^{-6}$
at a proton beam energy of 6 GeV/c. 
The only calculation  which reproduces this result, without violating the
constraints of low-energy data, is a
quark-model calculation of Preston and Goldman\cite{57},
in which parity violation occurs as a result of a quark-quark interaction in
an excited state of the proton which is formed as a result of an initial state
strong interaction.
This calculation  also predicts a rapid rise in $A_L$ as the beam
momentum increases.  Interest in this topic  was revitalized by the
possibility of new experiments involving polarized protons at RHIC and the AGS
at Brookhaven National Lab. T. Roser (BNL),
speaking at the June 1998 INT workshop
on parity violation, argued that it could be possible to ultimately make
measurements, with an error of about $3 \times 10^{-7}$,
at beam momenta between 5 and 250 GeV/c.
Working on understanding all of the reaction mechanisms is in progress  and
we have found that the physical ideas of Ref.~\cite{57} are very good, but that
the calculation is flawed in certain ways. Plans to make a better calculation
using a similar (but different) reaction mechanism are underway.

\vfill\eject

\subsection{Time Reversal and CP Violation}
{\it E. M. Henley}
\vskip 0.1in

Plans are underway to study time reversal and CP-violation for both heavy
quarks (the B system) and in nuclear physics.

\vskip 0.1in

\subsection{Deuteron Anapole Form Factor}
{\it M. J. Savage and 
R. P. Springer (Duke)}
\vskip 0.1in

Stimulated by the SAMPLE experiment at Bates, Springer and Savage
computed the anapole form factor of the deuteron at leading order with
effective field theory\cite{SSa}.  
Previously, they had computed the deuteron anapole
moment, and were able to extend this to the full form factor 
using the analytic
expressions for multi-loop integrals developed by 
Michael Binger\footnote{An undergraduate
student with the summer 1998 REU program at the UW}.
Due to the large size of the deuteron, it is possible, although very difficult,
to distinguish between
contributions from  
the nucleon strange form factor and 
from the deuteron anapole moment due 
to the different lengths scales involved in determining the scale of each form
factor.

\vskip 0.1in

\subsection{Parity Violation in $\gamma N$ Compton Scattering}
{\it P. F. Bedaque (INT) and  M. J. Savage }
\vskip 0.1in

It is now universally accepted that 
determining the single pion nucleon parity violating
coupling $h_{\pi NN}^{1}$ is very difficult.
Systems in which its effects are amplified do not allow for rigorous
computation
of observables, and systems where rigorous calculations are possible have only
small parity violating effects.
Experimental programs with ever-increasing sensitivity to parity violating
effects are ongoing in an effort to unambiguously determine  $h_{\pi NN}^{1}$.

In a somewhat speculative calculation,
Bedaque and Savage\cite{BSbs} examined parity violation in 
$\gamma N\rightarrow \gamma N$ Compton scattering.  
The amplitude for this process is
theoretically clean and is determined by the long sought after
parity violating $\pi NN$ coupling
$h_{\pi  NN}^{1}$.  Corrections to these contributions 
are small.
While measurement of such observables will be difficult, it would provide a
theoretically well-defined determination of $h_{\pi NN}^{1}$.

\vskip 0.1in
\subsection{Book on Symmetries}
{\it I. Halpern   and E. M. Henley}
\vskip 0.1in

The authors are preparing a book entitled
``{\bf\it Symmetries in Nature and Science}''
that is aimed at explaining
the importance of symmetries to a popular
audience. 

\vskip 0.2in
\section{Finite-Temperature QCD}
{\it J.O. Andersen (OSU), 
E. Braaten (OSU), 
and M. Strickland}
\vskip 0.1in

\begin{figure}
\epsfxsize = 11cm
\centerline{\epsfbox{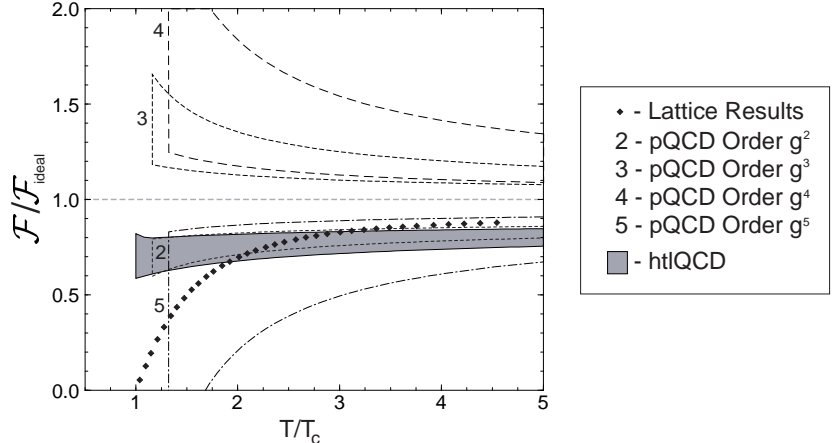}}
\caption[a]{The free energy of a hot gluon gas
        normalized to that of an ideal gas of
        gluons.
The black diamonds are  lattice data for pure-glue QCD~\cite{lattice}.
The bands enclosed by the curves labeled 2, 3, 4, and 5
      are the perturbative approximations to
      ${\cal F}_{QCD}$.
      The bands correspond to varying $\mu_4$ by a factor of 2.
The shaded region is the HTL-improved free energy.
      The region corresponds to varying the renormalization scale
      by a factor of 2.
}
\label{mikefig}
\end{figure}

The current theoretical understanding of the 
quark-gluon plasma (QGP) phase of nuclear matter
is rather limited.  In particular, the supposition that the QGP 
phase can be described as a weakly-interacting gas of quarks 
and gluons has lead to a number of theoretical predictions that rely 
on only the lowest order of perturbation theory.  In the last five 
years there have been some significant advances in finite-temperature 
perturbative calculations.   Of particular interest is the perturbative 
expansion of the free energy of the QGP phase.  This quantity has
been calculated to order $g^5$ \cite{qcd-pert,braaten-nieto} and 
show that 
in the region of experimental interest ($T < 1$ GeV) the series shows no 
signs of converging.  In Figure \ref{mikefig} the successive 
approximations to the free energy of a gluon plasma along with the latest 
lattice result are shown.  This clearly illustrates 
that the theoretical 
uncertainties due to the poor convergence of the perturbative series, 
even at order $g^5$, are quite large and that a different approach is needed.

The origin of the large perturbative corrections seems to be
plasma effects, such as the screening of interactions and the
generation of quasiparticles.  These effects
arise from the momentum scale $g T$.
Effective-field-theory methods can be used to isolate the effects of the
scale $g T$ from those of the scale $T$ \cite{braaten-nieto},
so that the effects of the scale
$g T$ can be calculated using nonperturbative methods.
There has been some recent 
progress in solving the analogous problem for a massless
scalar field theory with a $\phi^4$ interaction 
\cite{kpp,chiku-hatsuda}.  
When the free energy is calculated using
screened perturbation theory, the convergence of successive approximations
to the free energy is dramatically improved.  

\vskip 0.1in
\subsection{Scalar Field Theory - Screened Perturbation Theory}
\vskip 0.1in

At nonzero temperature, the conventional perturbative expansion 
of $g\phi^4$ theory
generates
infrared divergences. They can be removed by resumming the higher order
diagrams that generate a thermal mass of order $g T$ for the scalar particle.
This resummation changes the perturbative series from an expansion in powers
of $g^2$ to an expansion in powers of $g$.
Screened perturbation theory (SPT)
is simply a reorganization of the perturbation
series for thermal field theory.  
At nonzero temperature, SPT does not generate
any infrared divergences, because the mass parameter in the
free lagrangian provides an infrared cutoff.

Anderson, Braaten and Strickland 
have recently made a systematic three-loop calculation of the free energy,
entropy, and screening mass within SPT.  The previous calculation \cite{kpp} 
was 
only performed
to two loops because one of the three-loop diagrams (the so-called ``basketball
diagram'') had not been evaluated analytically.  
They have been able to 
calculate this diagram, reducing it to expressions 
involving three-dimensional integrals that can be easily evaluated 
numerically~\cite{bball}.  
enabling them to calculate the thermodynamic functions
for scalar field theories to three-loop order~\cite{bball,spt}.
The results of this calculation show that the convergence of the free-energy,
entropy, and screening mass are dramatically improved.

\vskip 0.1in
\subsection{Quantum Chromodynamics - Hard Thermal Loop Perturbation Theory}
\vskip 0.1in

In order to apply the techniques used in scalar theories to QCD
a fundamental change to the formalism must be made \cite{abs1}.
Instead of using a momentum-independent thermal mass one must use the 
momentum-dependent thermal gluon and quark masses which arise from 
resummation of hard-thermal-loop (HTL) diagrams.

Hard-thermal-loop perturbation theory (HTLpt) is a 
reorganization of the perturbation
series for thermal QCD.
In the last year Andersen, Braaten, 
and M. Strickland have completed the full one-loop HTLpt of the
QCD thermodynamic functions\cite{abs2}.  The shaded
band in Figure \ref{mikefig} is the result of the pure-glue calculation.  
The discrepancy between our result and the lattice result is due
the neglect of quasiparticle interactions which come in at two-loop
order.  These 
calculations demonstrate that HTLpt provides a tractable and gauge-invariant
method for  
incorporating the necessary physics.  
The two-loop calculation of the QCD free energy within HTLpt
is in progress \cite{abs3}.
When the two-loop calculation of the thermodynamic functions
is completed 
the same formalism can be used
to calculate real-time processes such as heavy quark and dilepton production
from a QGP.

\vskip 0.2in
\vfill\eject
\section{Many-Body Physics}
\vskip 0.1in

\subsection{Inhomogeneous Nuclear Matter} 
{\it A. Bulgac, P. Magierski (Warsaw),
and 
Y. Yu\footnote{PhD student of Bulgac.}
}
\vskip 0.1in

In Ref. \cite{ybm} it was shown that bubble fermion systems have a
remarkable soft collective branch, corresponding to the displacement
of the bubble. That study pointed to a large number of new physical
effects to be found in systems with bubbles. 
Bubble nuclei--even if
they exist--most likely would be almost impossible to
create in laboratory. 
However, similar effects could easily be observed in other
fermion systems, see Ref. \cite{bub2}. 
However, one of the most interesting
systems to consider are neutron stars. It was
predicted a long time ago that at about 0.5 km under the surface of a
neutron star nuclear matter is inhomogeneous and a sequence of bubble,
rod and plate phases should occur \cite{pet}. Almost all previous
studies of these phases of neutron matter have been performed within
the liquid-drop-model approximation.  One exception was the work of 
Refs. \cite{oya1,negele}, which however has a limited and incomplete
treatment of the shell-correction energy to ground-state energy of
the phase corresponding to nuclei embedded in a neutron gas. As we
discuss in Ref. \cite{bm} these last authors computed the least
important contribution to the ground state of inhomogeneous neutron
matter.

Until now nobody noticed a rather subtle
contribution to the ground state energy of inhomogeneous neutron
matter, for which there is naturally no established term yet in the
literature. One can term this contribution either as the Casimir
energy, as it is somewhat similar to the Casimir energy in quantum
field theory \cite{casimir}, or equally well use the more common term in
nuclear physics, shell correction energy.  Shell correction energy is
typically attributed to the difference in single particle spectra in
finite systems from a spectrum with a smooth level density. In
infinite matter however the spectrum is obviously continuous and one
might naively conclude that there is no shell correction contribution.

In order to better appreciate the nature of the problem we are
addressing in this work, let us consider the following simple situation. Let
us imagine that two spherical identical bubbles have been formed in an
otherwise homogeneous neutron matter. 
We shall ignore the role of long
range forces, namely the Coulomb interaction in the case of neutron
stars, as their main contribution is to the smooth, liquid drop or
Thomas--Fermi part of the total energy.
 
Under such circumstances one can ask the following apparently
innocuous question: ``What determines the most energetically favorable
arrangement of the two bubbles?'' According to a liquid drop model
approach (completely neglecting for the moment the possible
stabilizing role of the Coulomb forces) the energy of the system
should be insensitive to the relative positioning of the two
bubbles. A similar question was raised in condensed matter studies,
concerning the interaction between two impurities in an electron
gas. In the case of two ``weak'' and point--like impurities the
dependence of the energy of the system as a function of the relative
distance, ${\bf a}$, between the two impurities is given by 
the Ruderman--Kittel interaction:
\begin{equation}
E ({\bf a}) \propto \frac{\hbar ^2}{2 m k_{F}a^{3}} \cos (2 k_{F} a),
\end{equation}
where $k_{F}$ is the Fermi wave vector  and $m$ is the fermion mass.
In Ref. \cite{bm} we developed a method on how to compute the quantum
corrections to the ground state energy of inhomogeneous neutron
matter, based on quantum chaos techniques. Using the
Gutzwiller trace formula and taking into account only the shortest
classical period orbits, we have shown in particular that the
interaction energy between two isolated bubbles at large separations
($a=L-2R\gg R$)
\begin{equation}
E_{\circ \circ} \approx \frac{\hbar ^2 k_F^2}{2m} 
\left ( \frac{R}{a} \right ) ^2 \frac{2\sin (2k_F a)}{\pi ^2} .
\end{equation}
This result is unexpected in several respects, not the least its very
long-range interaction, c.f. the Ruderman-Kittel result above. 
We
quote explicitly only this result, as it displays the new qualitative nature
of our results. In Ref. \cite{bm} we have analyzed a variety of
inhomogeneous neutron phases: bubble phase, rod phase, plate phase,
local defects, various lattice deformations and temperature effects.
To our knowledge this is the first approach which considers
specifically the shell effects in the outside neutron gas and we aimed
at discussing its basic features.  Even though in principle
Hartree--Fock calculations include in principle such effects already,
the calculations performed so far \cite{negele} were too narrow in
scope and did not address this issue specifically and they were not able to
put in evidence a large number of new qualitative and quantitative issues.  
We have analyzed
the structure of the shell energy as a function of the density,
filling factor, lattice distortions and temperature.  The main lesson
we learned from this work is that the amplitude of the shell energy
effects is comparable with the energy differences between various
phases determined in simpler liquid drop type models. This fact alone
suggests that the inhomogeneous phase has most likely an extremely
complicated structure, maybe even completely disordered, with several
types of shapes present at the same time. It is clear that we have only managed
to ``barely scratch the surface'' of this problem.

Even though we have specifically addressed the case of neutron matter
only, it is clear that a similar picture should emerge when the
density increased and quarks become deconfined. In the low density
quark matter one would expect quark droplets of various shapes
embedded in a neutron gas \cite{pet}. 
It is highly likely that similar phenomena
are expected in the case of quark--gluon plasma and in the case of
strangelets, see Refs. \cite{madsen}. The calculation reported so far
in the literature
are very similar in spirit with the calculations performed in
Ref. \cite{oya1} for neutron matter, i.e. shell energy due to ``bound
fermion motion'' only. As we have shown however, most of the quantum
corrections to the ground state energy arises from the unbounded
fermion motion, which in a way can also be interpreted as the bubble--bubble
interaction energy, often mentioned in various papers but never evaluated,
since no methods have ever been developed in this direction. 
We also expect that
progress can be made in computing the Casimir energy as well for new 
geometries, both in QFT and in critical phenomena. We plan to address these 
range of questions in the near future.
Andreas Wirzba from Darmstadt TU has also expressed his enthusiasm and 
willingness to join our efforts.

\vskip 0.1in
\subsection{Dissipative Collective Motion }
{\it A. Bulgac, 
G. Do Dang (Orsay), and D. Kusnezov (Yale)}
\vskip 0.1in

We have continued to study the problem of quantum dissipative motion. 
Two review papers will shortly be published \cite{kbd}. 
In addition, a collaboration with Hans A. Weidenm\"uller has started. 
We are trying to implement new theoretical techniques  in
the derivation of the quantum kinetic equations based on the  
supersymmetric techniques, which are widely used in mesoscopic physics 
\cite{hans} and a new development, based on a generalization of the 
Keldysh--Schwinger formalism \cite{anton}. This new development is needed 
in order
to extend the applicability of our kinetic equations to lower temperatures
as well as in order to establish the limits of certain approximations used 
by us in our initial approach, in particular the role played by the so called
crossed (non--rainbow) diagrams in treating the intrinsic nuclear degrees of 
freedom. In a separate development with Do Dang and Kusnezov we have been able 
to extend significantly the calculation of the so called influence functional,
to the non--Markovian regime. 
The influence functional is the key element in 
this entire approach and is the object which describes the internal nuclear 
motion.

\vskip 0.1in

\subsection{Nuclear Energy Density Functional }
{\it A. Bulgac and V. Shaginyan (St. Petersburg)}
\vskip 0.1in

In the Fall of 1999 we finished writing an extensive computer
 program which 
describes spherical nuclei, using the new density functional due to Fayans
\cite{fayans} and
which also calculates the linear response to various external
fields. A number of theoretical 
developments were incorporated, which deal primarily with the
renormalization of the zero-range character of the residual interaction. Such 
modifications are required in order to evaluated correctly the Coulomb 
energy contributions to the nuclear ground state energies, 
when we have to evaluate 
loop integrals over the entire spectrum \cite{bsX,plb99}. 
We plan to perform
the actual calculations during the Fall 2000 INT program on Nuclear Structure.

\vskip 0.1in

\subsection{New Gaussian Single--Particle Wave Function Basis }
{\it A. Bulgac and Y. Yu}
\vskip 0.1in

We recently finished an initial study of a new
gaussian single--particle wave function basis to be used for the description
of the single--particle properties of many fermion systems, 
irrespective of the presence or absence of symmetries \cite{yy}. 
This new basis
consists of overlapping gaussians with a judicious choice of 
the distribution of
their centers, widths and number. We did not include the spin degrees of 
freedom in this initial study, due to some technical difficulties we have
encountered in dealing with them in an efficient manner.
In the absence of the spin--orbit
interaction we can describe spectra with an accuracy of about 0.02 MeV in the 
bulk and 0.15 MeV near the nucleon threshold.
The wave functions are reproduced
with an accuracy of not worse than a couple of percent, typically better.
The size of the basis set is around 
1000 basis functions for heavy nuclei. 
These results are significantly better than those obtained 
with other basis sets used in nuclear literature, e.g. harmonic oscillator wave
functions. We hope that with
more fine tuning we can improve both on the quality
and the number of the basis set and thus be able to propose this new basis 
as perhaps the basis set of choice for mean field calculations of deformed
nuclei.

\vskip 0.2in
\section{Hadronic Interactions and Structure}
\vskip 0.1in

\subsection{Strangeness Content of the Nucleon} 

{\it M. Alberg (Seattle U.), 
W. Hazelton\footnote{Former PhD student of Wilets and 
currently a Staff Scientist at the 
Fred Hutchinson Cancer Research Center.}, 
E. M. Henley, and L. Wilets.} 

\vskip 0.1in

Configurations containing
all $q^3$ and $q^4\bar q$ states, including $s$-$\bar s$ quark pairs, up to
a cut off of 1.5 GeV, and at most one gluon (OGE) have been included as
basis states in calculations of nucleons and hadrons.  Matrix
diagonalization, rather than perturbation theory, is employed in order to
allow for clustering of quark-antiquark pairs into meson-like structures ---
if such occur. In adjusting MIT parameters to fit energies, a reasonable {\it
positive} value for the Casimir term emerges, rather than the problematic
negative value employed in the original MIT model.  The results for
various observables are compared with experiment and with meson-based
models. Strange quarks give a minor contribution. Previous perturbation
calculations by two other groups were found to contain errors.

There is a delicate matter in the method:  The Hilbert space for 
diagonalization contains $q^3$ and $q^4\bar q$ but 
no explicit gluon states.  Gluons are contained  in effective
4-quark vertices.  The 
evaluation of the effective 
vertex involves the self-consistent energy of the state 
being evaluated.  The matrix diagonalization must be iterated to 
self-consistency.  Initial calculations used the unperturbed initial 3-quark 
energies.  Although the difference is expected to be small, work will 
continue.  The strange quark content of the nucleon was found to be quite 
small.

\vskip 0.1in

\subsection{Influence of Kaon-loops on $\phi$ Photoproduction}
{\it C. Hanhart, E. M. Henley, G.A. Miller,
and K. Nakayama (U. of Georgia)}
\vskip 0.1in

Questions about the strangeness content of the nucleon are widely discussed
in the literature. For some time $\phi$ 
photoproduction has been thought to be
a useful tool to study this. However, up to now there is no calculation 
available that studies the impact of $K$ and $K^*$ loops on this production.
We developed a formalism that will alow us to study this effect in a largely
model independent way. The results will be important in two respects: first
of all they will give insight into what one can really learn from 
$\phi$
photoproduction and should further constrain models for the strangeness
form factor of the nucleon.

\vfill\eject

\subsection{Production of Heavy Mesons in Nucleon-Nucleon Collisions}
{\it C. Hanhart, J. Speth (Julich), 
K. Nakayama (U. of Georgia), and J. Haidenbauer (Julich)}
\vskip 0.1 in

The project aims at an understanding of the physics
from polarization experiments of $\omega, \phi, \eta$ and
$\eta '$ production in nucleon nucleon collisions. Presently
we investigate approximation schemes to treat the nucleon-nucleon interaction
in the initial state, 
for there is, as yet, 
no model of NN scattering at these high energies.
Since 
protons probes provide complementary information to electro-magnetic
probes, we hope meson production in nucleon nucleon collisions, when
studied consistently with electro production of heavy mesons, allows
for deeper insight into the phenomenology of missing resonances.

\vskip 0.1in 
\subsection{A Meson-Exchange Model for Pion 
Production in Nucleon-Nucleon Collisions}
{\it C. Hanhart,  J. Speth (Julich), 
C. Elster (Ohio), and J. Haidenbauer (Julich)}
\vskip 0.1in 

We have developed a meson-exchange model for pion production in 
nucleon-nucleon
collisions\cite{Hana}
 that has proven very successful in the description of charged
pion production in both unpolarized and polarized observables. However,
in case of the $\pi^0$ production we still see discrepancies that are not
fully understood. So far the final state was treated as an
effective two body problem by means of a distorted wave approximation.
We are presently working on an improvement of the 
three body treatment of the problem in order to resolve the discrepancy 
between theory and experiment.

\vskip 0.1 in

\subsection{$p p\rightarrow p \Lambda K^+$ and 
$p p \rightarrow p \Sigma ^0 K^+$ Near Threshold}
{\it A.~Gasparian (Moscow), 
J.~Haidenbauer (Julich), 
C.~Hanhart, L.~Kondratyuk (Moscow),  and 
J.~Speth (Julich)}
\vskip 0.1 in

Very little is known about the hyperon--nucleon interaction at low energies. 
One possible way to
gain experimental information on this system is through production reactions. 
At low energies there
are two interesting quantities accessible: 
the energy dependence allows a determination of low energy
hyperon--nucleon scattering parameters and the 
total strength of the cross section, that contains
information on both the hyperon nucleon interaction 
as well as the production operator. 
Our work\cite{GHHKS}
shows that the energy dependence of the total cross sections 
is in agreement with predictions
from the existing hyperon-nucleon potentials. 
In order to explain the total strength of the 
cross section a destructive interference 
between pion and kaon exchange is required.
The relative sign between these exchanges is the only 
free parameter of the model. This allowed
Hanhart and collaborators 
to propose the measurement of the reaction 
$pp\to n\Sigma^+K^+$ to check the model. Their investigation
was the first microscopic investigation of the associated 
strangeness production at close to the
threshold energies. 
Further studies aim at a better treatment of the meson-baryon interactions. 

\vskip 0.1in

\subsection{The Structure of the Roper Resonance}
{\it O.~Krehl (Julich), 
C.~Hanhart, 
S.~Krewald (Julich) 
and J.~Speth (Julich)}
\vskip 0.1in

The spectrum of baryon excitations is still not fully understood.
There are different
mechanisms discussed in the literature 
to understand why the Roper resonance, 
the lightest positive 
parity excitation, 
has a lower mass than the lightest negative parity excitation. 
If the binding
where dominated by the confining potential the inverse  order 
is expected. 
Hanhart and collaborators\cite{Roper} 
took a different point of view to quark models: 
since most of the experimental information
on the Roper comes from pion induced reactions their goal 
was to  describe elastic and
inelastic pion nucleon scattering. 
By employing a coupled-channel meson-exchange model they where
able to describe the $\pi N$ scattering phase shifts as well 
as inelasticities up to center of
mass energies of $1.9~{\rm GeV}$. 
A good description of the Roper partial wave $P_{11}$ was
possible without the inclusion of a bare pole diagram. 
This investigation demonstrates that there
is not necessarily a problem with the baryon spectrum 
since the Roper resonance might  be
generated dynamically. 
In any case: we demonstrated that the meson baryon interaction in the 
$P_{11}$ partial wave is too strong to be 
neglected when trying to understand the baryon spectrum.

\vskip 0.1in
\subsection {Pion-Baryon Couplings and SU(3)}

{\it A. Buchmann (Tubingen) and E. M. Henley }
\vskip 0.1in

Work has been completed\cite{Buchmann:1999ab}
on extending a  method devised by Morpugo to obtain
static properties of nucleons. This method has been used to calculate and
predict
meson-baryon couplings to the octet and decuplet baryons. 
This work has been submitted for publication. A talk on the subject at the
3d Annual Symposium on Symmetries in Subatomic Physics will also be published.

This method has also been used to predict quadrupole moments and quadrupole
transition moments for the decuplet baryons and between the octet and decuplet.
This work is almost ready to submit for publication.

\subsection{Electron-Deuteron Scattering in a Relativistic 
Current-Conserving Framework} 

{\it D.~R.~Phillips and 
S.~J.~Wallace (Maryland)}
\vskip 0.1in

The inclusion of relativity in theoretical models becomes particularly
critical in the description of electron-deuteron scattering
experiments. In electron-deuteron scattering the standard
non-relativistic quantum mechanical treatment of the problem is
expected to fail at energies where interesting physics occurs. The
various available relativistic formalisms all produce different
predictions for the observables, none of which accurately describes
all of the experimental data. 
Phillips and Wallace have 
developed an alternative three-dimensional relativistic scheme for
calculating these processes~\cite{ourwork}. Impulse approximation
calculations of the experimentally-measured quantities $A$, $B$, and
$T_{20}$ show good agreement for $T_{20}$ out to 
$Q^2 \sim 1 \mbox{GeV}^2$. 
This calculation is missing strength in $A$ and
$B$~\cite{Nealpaper}, while approaches which are essentially
non-relativistic reproduce the data quite well~\cite{AV18paper}. From
this one might conclude that the non-relativistic approach is the
correct one. Yet it is clear that relativity should begin to play a
crucial role in the description of data at these $Q^2$'s.

In December Phillips visited the Thomas Jefferson National Accelerator
Facility and spent three weeks working there with 
Wallace on these issues. The time was productive, and they have  made
much progress in improving the framework they have been working on
together for the past four years. In particular, the connection of this
work to the usual non-relativistic approaches to electron-deuteron
scattering is now much clearer. Further calculations will be pursued
in the coming year.

\vskip 0.2in
\section{Large-$N$ QCD and Nuclei}

{\it S.~R. Beane}
\vskip 0.1in

The large-$N$ approximation ---where $N$ is the number of 
colors--- is one of the few systematic approaches to QCD at
low-energies~\cite{'tHooft}\cite{Witten}. The large-$N$ approximation
yields a great deal of predictive power, particularly in the baryon
sector. Less is known in the meson sector. For instance, according to
canonical large-$N$ wisdom, meson-meson scattering is governed by
exchange of an infinite number of mesons in tree graph
approximation. However, little progress has been made in utilizing
this property to make predictions.

\vskip 0.1in

\subsection{High Energy Theorems at Large-$N$}
\vskip 0.1in

In recent work it was shown that spectral function sum
rules~\cite{Weinberg} take a particularly simple algebraic form in the
large-$N$ limit~\cite{Knecht}. This allows one to correlate the
ordering pattern of narrow meson states at low energies with the size
of local order parameters of chiral symmetry breaking in QCD. The
spectral function sum rules are constraints on time ordered products
of two QCD currents and traditionally their derivation relies on the
technology of the operator product expansion.

Recently Beane showed that in the large-$N$ limit one can derive sum rules
for products of two, three and four QCD currents using chiral symmetry
at infinite momentum in the large-$N$ limit~\cite{HET1}.  These exact
relations among meson decay constants, axialvector couplings and
masses determine the asymptotic behavior of an infinite number of QCD
correlators. The familiar spectral function sum rules for products of
two QCD currents are among the relations derived.  With this precise
knowledge of asymptotic behavior, an infinite number of large-$N$ QCD
correlators can be constructed using dispersion relations. Beane gave a
detailed derivation of the exact large-$N$ pion vector form factor and
forward pion-pion scattering amplitudes.

\vskip 0.1in

\subsection{Low Energy Constants from High Energy Theorems}
\vskip 0.1in

As a phenomenological application of the high-energy theorems, 
Beane
considered new constraints implied by the sum rules on resonance
saturation in chiral perturbation theory~\cite{HET2}. The sum rules
imply that the low-energy constants of chiral perturbation theory are
related by a set of mixing angles. Beane showed that the simplest
nontrivial saturation scheme predicts low-energy constants that are in
remarkable agreement with experiment. Further, it was shown that (1)
vector-meson dominance can be understood as a consequence of the fact
that nature has chosen the lowest-dimensional nontrivial chiral
representation, and (2) chiral symmetry places an upper bound on the
mass of the lightest scalar in the hadron spectrum.

\vskip 0.2in
\section{Other}
\vskip 0.1in

\vskip 0.1in

\subsection{Dimensional Regularization of Integral Equations}

{\it D.~R.~Phillips, 
I.~R.~Afnan (Flinders), 
and A.~G.~Henry-Edwards (Flinders)
}
\vskip 0.1in

One difficulty for standard treatments of hadronic reactions is that
form factors are always used to regulate integrals which would otherwise
be divergent. This procedure is well-motivated, since we know that
the hadrons have substructure which gives them a finite extent, and hence,
a form factor. However, if we understand this description of hadronic 
dynamics as a field theory then it is non-local, and hence the implementation
of basic field theoretic principles, such as gauge invariance, can be quite
involved. Methods have been formulated to impose gauge invariance 
on an amplitude containing hadronic form factors, but they are intrinsically
non-unique, since they constrain only the longitudinal part of the
electromagnetic amplitude.

Some of these difficulties can be resolved by using dimensional
regularization (DR) to render divergent integrals finite. This is the
method of choice for dealing with the infinities which arise in
perturbative field-theoretic calculations.  However, there are very
few studies of the application of dimensional regularization to
integral equations, as opposed to its many applications to integrals
in perturbative calculations. 
Phillips, Afnan and Henry-Edwards\cite{PAHE}
recently implemented the
numerical solution of this dimensionally-regulated integral equation,
and have extracted physical quantities, such as phase shifts, once the
regulation and renormalizaton have been done.

\vskip 0.1in
\subsection{Exact Solutions to the Schroedinger Equation for
Coulomb-Plus-Oscillator Potentials}
{\it M. Alberg (Seattle U.) and L. Wilets}
\vskip 0.1in
        
In our study of the infamous cold fusion problem (which yielded a 
bound on the process orders of magnitude less than claimed), we 
calculated wave functions of two nuclei in a plasma.  We
discovered an infinite set of exact solutions to the Coulomb plus harmonic
oscillator potentials.  Each solution, corresponds to a 
particular ratio of the coefficients of the two terms, and hence the set is 
not orthogonal.  There is considerable interest in special soluble 
potentials.  A particular application is in the numerical calculation of 
wave functions for potentials containing singular terms, such
as the Coulomb term.
If one writes the radial function as $u(r)=u_0(r)\,f(r)$, where $u_0$ 
is an exact solution for a potential with the same singularity, then 
$f (r)$ is very 
well-behaved, especially if one chooses a solution of similar eigenvalue.
A paper is in final preparation.


\begin{thebibliography}{99}
\bibitem{Wein} S. Weinberg, Phys. Lett. B {\bf 251}, 288 (1990); 
  Nucl. Phys. {\bf B363}, 3 (1991); Phys. Lett. B {\bf 295},
  114 (1992).

\bibitem{BNTTa} S. Burles, K. M. Nollet, J. W. Truran and M. S. Turner,
Phys. Rev. Lett. {\bf 82}, 4176 (1999).

\bibitem{CSa} J.-W. Chen and M. J. Savage,
Phys. Rev.  {\bf C60}, 065205 (1999). 

\bibitem{Ra} G. Rupak, {\tt nucl-th/9911018},
to appear in Nucl. Phys. {\bf A}.

\bibitem{Compton} S.~R.~Beane, M.~Malheiro, D.~R.~Phillips,
and U.~van Kolck, Nucl. Phys. {\bf A656}, 367 (1999).

\bibitem{Lucas} M.~Lucas, Ph. D. thesis, Univ. of Illinois (1994).

\bibitem{Hornidge} D.~Hornidge {\it et al.}, Phys. Rev. Lett.
{\bf 84}, 2334 (2000).

\bibitem{nadia} N. Fettes, and U.-G. Mei\ss ner, hep-ph/0002162.

\bibitem{CRSnopi} J. -W. Chen, G. Rupak and M. J. Savage,
Nucl. Phys. {\bf A653},  386  (1999).

\bibitem{PotSch}
R. Schiavilla {\it et al}.,
Phys. Rev. {\bf C58}, 1263 (1998).

\bibitem{BCa} M. N. Butler and J.-W Chen,
{\tt nucl-th/9905059}.

\bibitem{BCb}M. N. Butler and J.-W Chen, 
In preparation.

\bibitem{YHH} S. Ying, W. C. Haxton and E. M. Henley,
Phys. Rev. {\bf C45}, 1982 (1992);
Phys. Rev. {\bf D40}, 3211 (1989).

\bibitem{KN} K. Kubodera and S. Nozawa,
Int. J. Mod. Phys. {\bf E3}, 101 (1994);
Y. Kohyama and K. Kubodera, USC(NT)-report-92-1, unpublished.

\bibitem{BGGa} F. Gabbiani, P. F. Bedaque and H. W. Grie\ss hammer,
{\tt nucl-th/9911034}.

\bibitem{HMKa}
C. Hanhart, G. A. Miller and U. van Kolck,
{\tt nucl-th/0004033}.

\bibitem{24} G.A.~Miller,
Phys. Rev. {\bf C56}, 8 (1997), Phys. Rev. {\bf C56}, 2789 (1997).

\bibitem{25}G.A.~Miller
and R.~Machleidt, Phys.\ Lett.\  {\bf B455}, 19 (1999).

\bibitem{26} 
M.~Burkardt and G.A.~Miller,
Phys. Rev. {\bf C58}, 2450 (1998)
nucl-th/9802049.

\bibitem {27}
P.G.~Blunden, M.~Burkardt and G.A.~Miller,
Phys.\ Rev.\  {\bf C59}, R2998 (1999);
Phys.\ Rev.\  {\bf C60}, 055211 (1999)

\bibitem {28} 
G.A.~Miller and R.~Machleidt, 
Phys.\ Rev.\  {\bf C60}, 035202 (1999)

\bibitem{review}  G.~A.~Miller,
``Light front
quantization: A technique for relativistic and realistic  nuclear physics,''
Accepted by Prog. Part. Nucl. Phys. nucl-th/0002059.

\bibitem{QMC} P.A.M.\ Guichon, Phys.\ Lett.\ B {\bf 200}, 235 (1988).
K.\ Saito and A.W.\ Thomas, Phys.\ Lett.\ B {\bf
327}, 9 (1994);
P.G.~Blunden and G.A.~Miller,
Phys. Rev. {\bf C54}, 359 (1996).

\bibitem{15} L. Frankfurt, G.A. Miller and M. Strikman, 
Phys. Lett. B304, 1  (1993).

\bibitem{Ackerstaff:2000ac}
K.~Ackerstaff {\it et al.}  [HERMES Collaboration],
Phys.\ Lett.\  {\bf B475}, 386 (2000)

\bibitem{Callan:1969uq}
C.~G.~Callan and D.~J.~Gross,
Phys.\ Rev.\ Lett.\  {\bf 22}, 156 (1969).

\bibitem{Miller:2000ta}
G.~A.~Miller, S.~J.~Brodsky and M.~Karliner,
Phys.\ Lett.\  {\bf B481}, 245 (2000)

\bibitem{Blunden:2000wb}
P.~G.~Blunden, M.~Burkardt and G.~A.~Miller,
Phys.\ Rev.\  {\bf C61}, 025206 (2000)

\bibitem{RMa}
R. Michaels, talk at HAPPEX collaboration meeting, TJNAF,
Dec. (1998).

\bibitem{HPSM}
 C. J. Horowitz, S. J. Pollock, P. A. Souder,
R. Michaels, {\tt nucl-th/9912038}.

\bibitem{Cooke:2000yi}
J.~R.~Cooke, G.~A.~Miller and D.~R.~Phillips,
Phys.\ Rev.\  {\bf C61}, 064005 (2000)

\bibitem{Cooke:2000ef}
J.~R.~Cooke and G.~A.~Miller,
nucl-th/0002016.



\bibitem{29} 
 L. Frankfurt and
M. Strikman, Physics Reports {160}, 237 (1988);
L.L.~Frankfurt, G.A.~Miller and M.~Strikman,
Ann. Rev. Nucl. Part. Sci. {\bf 44}, 501 (1994).


\bibitem{35}  G.A. Miller, B.M.K. Nefkens and I. Slaus,
Phys. Rpts. {194},1 (1990);
G.A. Miller and W.T.H. Van Oers, p.127 in ``Symmetries and
Fundamental Interactions in Nuclei" edited by W.C. Haxton and E.M. Henley,
World Scientific (Singapore) 1995; 



\bibitem{38}
H.O. Meyer {\it et al.}, Nucl. Phys. {\bf A539}, 633 (1992);
A. Bondar {\it et al.}, Phys. Lett. {\bf B356}, 8 (1995).

\bibitem{39} 
T.D.~Cohen, J.L.~Friar, G.A.~Miller and U.~van Kolck,
Phys. Rev. {\bf C53}, 2661 (1996).



\bibitem{41}
T.~Sato, T.S.~Lee, F.~Myhrer and K.~Kubodera,
Phys. Rev. {\bf C56}, 1246 (1997).

\bibitem{42} 
I.~Sick and D.~Day,
Phys. Lett. {\bf B274}, 16 (1992).

\bibitem{46}
T.~Nieuwenhuis and J.A.~Tjon,
Phys. Rev. Lett. {\bf 77}, 814 (1996).

\bibitem{50} 
E.A.~Hawker {\it et al.}
Phys. Rev. Lett. {\bf 80}, 3715 (1998);
J.C.~Peng {\it et al.}
Phys. Rev. {\bf D58}, 092004 (1998).

\bibitem{Alberg:2000bc}
M.~Alberg, E.~M.~Henley and G.~A.~Miller,
Phys.\ Lett.\  {\bf B471}, 396 (2000).

\medskip
\bibitem{52} D. Ashery, 
hep-ex/9910024.
\medskip
\bibitem{Frankfurt:1999tq}
L.~Frankfurt, G.~A.~Miller and M.~Strikman,
{\tt hep-ph/9907214}, 
in press Foundations of Physics.

\bibitem{53}C.~Bochin {\it et al.}, Phys. Rev. Lett. {\bf 81}, 4576 (1998).

\bibitem{Frankfurt:2000ik}
L.~L.~Frankfurt, G.~A.~Miller, M.~M.~Sargsian and M.~I.~Strikman,
Phys.\ Rev.\ Lett.\  {\bf 84}, 3045 (2000)

\bibitem{MF} B.~L.~Friman and O.~V.~Maxwell, Ap. J. {\bf 232}, 541 (1979).

\bibitem{WWRa}M. J. Watrous, L. Wilets and J. J. Rehr, 
Phys. Rev. {\bf E59}, 3554 (1999).

\bibitem{WGWRa} 
L. Wilets, B. Giraud,  M. J. Watrous and J.J. Rehr, 
Ap. J. {\bf 530}, 504 (2000).  


\bibitem{alford1}
M.~Alford, K.~Rajagopal and F.~Wilczek,
Phys. Lett. {\bf B422}, 247 (1998);
R.~Rapp, T.~Sch\"afer, E.V.~Shuryak and M.~Velkovsky,
Phys. Rev. Lett. {\bf 81}, 53 (1998).

\bibitem{BBSa} S. R. Beane, P. F. Bedaque and M. J. Savage,
Phys. Lett. {\bf B483}, 131 (2000).

\bibitem{BBSb} S. R. Beane, P. F. Bedaque and M. J. Savage,
{\tt nucl-th/0004013}.

\bibitem{BBc} P. F. Bedaque and S. R. Beane,
{\tt nucl-th/0005052}.


\bibitem{ivwei1} 
S. Weinberg, Trans. N.Y. Acad. Sci. {\bf 38} (1977) 185.
\bibitem{VNM00} U. van Kolck,
J.A. Niskanen, and G.A.~Miller, submitted to PLB.
nucl-th/0006045 


\bibitem {56} 
N.~Lockyer {\it et al.},
Phys. Rev. Lett. {\bf 45}, 1821 (1980);  
Phys. Rev. {\bf D30}, 860 (1984).

\bibitem{57}
T.~Goldman and D.~Preston,
Phys. Lett. {\bf 168B}, 415 (1986).

\bibitem{SSa}M. J. Savage and R. P. Springer,
{\tt nucl-th/9907069}.

\bibitem{BSbs} P. F. Bedaque and M. J. Savage,
{\tt nucl-th/9909055}, to appear in Phys. Rev. {\bf C}.


\bibitem{qcd-pert}
P. Arnold and C. Zhai, Phys. Rev. {\bf D50}, 7603 (1994);
        Phys. Rev. {\bf D51}, 1906 (1995); B.~Kastening and C.~Zhai, 
        Phys. Rev. {\bf D52}, 7232 (1995).

\bibitem{braaten-nieto}
E.~Braaten and A.~Nieto, Phys. Rev. Lett. {\bf 76}, 1417 (1996);
        Phys. Rev. {\bf D53}, 3421 (1996).

\bibitem{lattice}
G. Boyd et al., Phys. Rev. Lett. {\bf 75}, 4169 (1995);
        Nucl. Phys. {\bf B469}, 419 (1996);

\bibitem{kpp}
F. Karsch, A. Patk\'os, and P. Petreczky, Phys. Lett. {\bf B401}, 69 (1997).

\bibitem{chiku-hatsuda}
S. Chiku and T. Hatsuda, Phys. Rev. {\bf D58}, 076001 (1998);
        hep-ph/9809215.

\bibitem{bball}
J.O. Andersen, E. Braaten, and M. Strickland, hep-ph/0002048,
accepted for publication in Physical Review D.

\bibitem{spt}
J.O. Andersen, E. Braaten, and M. Strickland, 
Screened Perturbation Theory to Three Loops,
in preparation.

\bibitem{abs1} 
J.O. Andersen, E. Braaten and M. Strickland, 
        Phys. Rev. Lett. {\bf 83}, 2139 (1999);
        Phys. Rev. {\bf D61}, 14017 (2000).

\bibitem{abs2}
J.O. Andersen, E. Braaten and M. Strickland, Phys. Rev. {\bf D61}, 74016 (2000).

\bibitem{abs3}
J.O. Andersen, E. Braaten, and M. Strickland, 
HTL Perturbation Theory to Two Loops,
in preparation.


\bibitem{ybm} Y. Yu, A. Bulgac and P. Magierski, 
Phys. Rev. Let. {\bf 84}, 412 (2000).


\bibitem{bub2} A. Bulgac {\it et al.},  
in {\it Proc. Intern. Work. on
Collective excitations in Fermi and Bose systems}, 
eds. C.A. Bertulani
and M.S. Hussein (World Scientific, Singapore 1999), pp 44--61.


\bibitem{pet}  G. Baym {\it et al.}, Nucl. Phys.
{\bf A175}, 225 (1971); C.J. Pethick and D.G. Ravenhall,
Annu. Rev. Nucl. Part. Sci.  {\bf 45}, 429 (1995); H. Heiselberg {\it
et al.}, Phys. Rev. Lett.  {\bf 70}, 1355 (1992).


\bibitem{oya1} K. Oyamatsu, Nucl. Phys. {\bf A561}, 431 (1993), 
K. Oyamatsu, M. Yamada, Nucl. Phys. {\bf A578}, 181 (1994).


\bibitem{negele} J.W. Negele and D. Vautherin, Nucl. Phys. {\bf A207}, 
  298 (1973); P. Bonche and D. Vautherin, Nucl. Phys. {\bf A372}, 496
  (1981); Astron. Astrophys. {\bf 112}, 268 (1982).


\bibitem{bm} A. Bulgac and P. Magierski, astro--ph/0002377.


\bibitem{casimir} M. Kardar and R. Golestanian, Rev. Mod. Phys. {\bf
    71}, 1233 (1999) and references therein.


\bibitem{madsen} G. Neegaard and J. Madsen, hep--ph/0003176;  J. Madsen, 
astro--ph/9809032.


\bibitem{kbd} D. Kusnezov, A. Bulgac and G. Do Dang, nucl-th/9911069;
A. Bulgac, G. Do Dang and D. Kusnezov, quant-ph/9911098; both to be published 
in Physica {\bf A}.


\bibitem{hans} J.J.M. Verbaarschot. M. Zirnbauer and H.A. Weidenm\"uller, 
Phys. Rep. {\bf 6}, 367 (1985); 
K. Efetov, {\it Supersymmetry}.  


\bibitem{anton} A. Kamenev and A. Andreev, Phys. Rev. B {\bf 60} 2218 (1999);
C. Chamon, A.W.W. Ludwig and C. Nayak, Phys. Rev. B {\bf 60}, 2239 (1999).


\bibitem{fayans} S.A. Fayans, JETP Letters, 68 (1998) 169.



\bibitem{bsX} A. Bulgac and V.R. Shaginyan, 
to be published in {\it Advances in Quantum
Many-Body Theory}, vol. 3, eds. R.F. Bishop {\it et al.}
(World Scientific, Singapore). 

\bibitem{plb99} A. Bulgac and V.R. Shaginyan, 
Phys. Lett. {\bf B469} 1 (1999). 

\bibitem{yy} Y. Yu and A. Bulgac, to be submitted.

\bibitem{Hana} C. Hanhart,
Invited talk at the 
4th International Conference on Physics at Storage Rings (STORI99), 
Bloomington, Sept. 1999,

\bibitem{GHHKS}
 A. Gasparian, J. Haidenbauer, C. Hanhart,
L. Kondratyuk, and  J. Speth
Phys. Lett. {\bf B480}, 273 (2000).

\bibitem{Roper}
O. Krehl, C. Hanhart, S. Krewald, J. Speth,
{\tt nucl-th/9911080 }.


\bibitem{Buchmann:1999ab}
A.~J.~Buchmann and E.~M.~Henley,
{\tt nucl-th/9912044}.

\bibitem{'tHooft}
G.~'t Hooft,
Nucl.\ Phys.\  {\bf B72}, 461 (1974).

\bibitem{Witten}
E.~Witten,
Nucl.\ Phys.\  {\bf B156}, 269 (1979).

\bibitem{Weinberg}
S.~Weinberg,
Phys.\ Rev.\ Lett.\  {\bf 18}, 507 (1967).

\bibitem{Knecht}
M.~Knecht and E.~de Rafael,
Phys.\ Lett.\  {\bf B424}, 335 (1998).

\bibitem{HET1} S.~R.~Beane,
Phys.Rev. {\bf D61}, 116005 (2000).

\bibitem{HET2} S.~R.~Beane,
To appear in Journal of Physics G: Nuclear and Particle Physics.


\bibitem{ourwork}  D.~R.~Phillips and S.~J.~Wallace, Phys. Rev. C
{\bf 54}, 507 (1996);  Few Body Syst. {\bf 24}, 175 (1998).

\bibitem{Nealpaper} D.~R.~Phillips, N.~K.~Devine, and
S.~J.~Wallace, Phys. Rev. C {\bf 58}, 2261 (1998).

\bibitem{AV18paper} R.~B.~Wiringa, V.~G.~J.~Stoks, and R.~Schiavilla,
Phys.Rev.C {\bf 51}, 38 (1995).

\bibitem{PAHE} D.R. Phillips, I.R. Afnan, and A.G. Henry-Edwards,
Phys. Rev. {\bf C61}, 044002 (2000). 

\end{thebibliography}
\end{document}